\documentclass[a4paper]{spie}  
\usepackage[]{graphicx}
\usepackage{amssymb}
\usepackage{array}
\usepackage{url}

\title{A coherent polarimeter array for the Large Scale Polarization Explorer balloon experiment} 

\author{
M. Bersanelli\supit{1,2}, 
A. Mennella\supit{1,2}, 
G. Morgante\supit{3}, 
M. Zannoni\supit{4}, 
G. Addamo\supit{5}, 
A. Baschirotto\supit{4}, 
P. Battaglia\supit{1}, 
A. Ba\`u\supit{4}, 
B. Cappellini\supit{1,2}, 
F. Cavaliere\supit{1}, 
F. Cuttaia\supit{3}, 
F. Del Torto\supit{1}, 
S. Donzelli\supit{2}, 
Z. Farooqui\supit{5}, 
M. Frailis\supit{6}, 
C. Franceschet\supit{1},
 E. Franceschi\supit{3},
 T. Gaier\supit{7}, 
S. Galeotta\supit{6}, 
M. Gervasi\supit{4}, 
A. Gregorio\supit{8,6}, 
P. Kangaslahti\supit{7}, 
N. Krachmalnicoff\supit{1}, 
C. Lawrence\supit{7}, 
G. Maggio\supit{6}, 
R. Mainini\supit{4}, 
D. Maino\supit{1,2}, 
N. Mandolesi\supit{3}, 
B. Paroli\supit{1}, 
A. Passerini\supit{4}, 
O. A. Peverini\supit{5}, 
S. Poli\supit{1}, 
S. Ricciardi\supit{3}, 
M. Rossetti\supit{1}, 
M. Sandri\supit{3}, 
M. Seiffert\supit{7}, 
L. Stringhetti\supit{2}, 
A. Tartari\supit{4}, 
R. Tascone\supit{5}, 
D. Tavagnacco\supit{8,6}, 
L. Terenzi\supit{3}, 
M. Tomasi\supit{1,2}, 
E. Tommasi\supit{9}, 
F. Villa\supit{3}, 
G. Virone\supit{5}, 
A. Zacchei\supit{6}
\skiplinehalf
\supit{1}Dipartimento di Fisica, Universit\`a degli Studi di Milano, via Celoria, 16, 20133 Milano, Italy;\\
\supit{2}INAF-IASF Milano, via E. Bassini 15, 20133 Milano, Italy;\\
\supit{3}INAF-IASF Bologna, via Gobetti 101, 40129 Bologna, Italy;\\
\supit{4}Dipartimento di Fisica,  Universit\`a di Milano - Bicocca, P.zza della Scienza 3, 20126 Milano, Italy;\\
\supit{5}CNR-IEIIT, c/o Politecnico di Torino, Corso Duca degli Abruzzi 24, 10129 Torino, Italy;\\
\supit{6}INAF Ð Osservatorio Astronomico di Trieste, via G.B. Tiepolo 11, 34131 Trieste, Italy;\\
\supit{7}Jet Propulsion Laboratory, California Institute of Technology, 4800 Oak Grove Drive, Pasadena, California, USA;\\
\supit{8}Dipartimento di Fisica, Universit\`a degli Studi di Trieste, via A. Valerio 2, 34127 Trieste, Italy;\\
\supit{9}Agenzia Spaziale Italiana, viale Liegi 26, 00198 Roma, Italy
}

\authorinfo{Further author information: (Send correspondence to Marco Bersanelli)\\Marco Bersanelli.: E-mail: marco.bersanelli@mi.infn.it, Telephone: 39 02 50317264}

\begin{document} 
  \maketitle 

\begin{abstract}
We discuss the design and expected performance of STRIP (STRatospheric Italian Polarimeter), an array of coherent receivers designed to fly on board the LSPE (Large Scale Polarization Explorer) balloon experiment. The STRIP focal plane array comprises 49 elements in Q band and 7 elements in W-band using cryogenic HEMT low noise amplifiers and high performance waveguide components. In operation, the array will be cooled to 20 K and placed in the focal plane of a $\sim 0.6$ meter telescope providing an angular resolution of  $\sim1.5$ degrees. The LSPE experiment aims at large scale, high sensitivity measurements of CMB polarization, with multi-frequency deep measurements to optimize component separation. The STRIP Q-band channel is crucial to accurately measure and remove the synchrotron polarized component, while the W-band channel, together with a bolometric channel at the same frequency, provides a crucial cross-check for systematic effects.
\end{abstract}



\section{INTRODUCTION}
\label{sec:intro}  

  Since its discovery in 1965, the Cosmic Microwave Background (CMB) has offered fundamental cosmological information at every level of precision that new experiments have been able to probe. The next ``accuracy standards'' in the field will be set by the cosmological results of the Planck space mission\cite{Planck_01}, to be released in 2013 and 2014, which are expected to yield high-precision measurements of the CMB temperature anisotropies and to significantly improve over the current status of polarization measurements. However, much will be left after Planck for CMB polarization studies. The challenge for the next generation of experiments is to measure the polarization properties of the CMB to sub-micro-Kelvin level over a large portion of the sky. In fact, a high-precision reconstruction of the CMB polarization power spectrum at low multipoles (below $l\sim 100$) represents the best opportunity for a direct measurement of the B-mode polarized component, whose detection would unveil an explicit signature of primordial gravitational waves from an inflationary era in the very early universe. The search for B-mode polarization requires technologically advanced instruments using large arrays of detectors. Extreme sensitivity, unprecedented control of systematic effects ($<1$ $\mu$K), as well as a wide spectral range to disentangle polarized foreground emission are the key requirements.

The LSPE (Large Scale Polarization Explorer) balloon program is designed to contribute to this new challenge. The experiment\cite{DeBernardis01} covers the frequency range 40 - 220 GHz in 4 frequency bands with two instruments, one  based on coherent polarimeters (STRIP, STRatospheric Italian Polarimeter) and the other on bolometric detectors (SWIPE, Short Wavelength Instrument for the Polarization Explorer\cite{SWIPE01}). The LSPE program, supported by the Italian Space Agency (ASI), aims at both competitive scientific results and technological development in the field.

The STRIP instrument covers the low frequency side of the LSPE range with 49 polarimeter modules in Q band ($\sim 40$ GHz) and a small array of 7 modules in W band ($\sim 90$ GHz). The Q band channel is designed to map to exquisite precision the polarized Galactic synchrotron component, which is dominant in a relatively large fraction of the sky, and to contribute to the CMB polarized sensitivity in foreground-clean regions. The W band array, in conjunction with a bolometric channel in the same frequency band, provides a key leverage for systematic effect crosscheck by exploiting the widely different technologies adopted by the two instruments. The STRIP polarimeter modules are based on the design of the QUIET receivers\cite{quiet_2011}, which provides excellent redundancy and suppression of systematic effects inherent to the receivers. The Radio Frequency (RF) chain is based on cryogenic  High Electron Mobility Transistor (HEMT) low noise amplifiers and on high-performance waveguide components cooled to 20\,K and integrated in Monolithic Microwave Integrated Circuits (MMIC). The instrument will be placed in the focal plane of a dedicated dual-reflector telescope providing an angular resolution of about 1.5 degrees. In this paper we review the scientific requirements and present the overall configuration and main design solutions adopted for the LSPE/STRIP instrument. 

\section{SCIENCE CASE}
 \label{sec:2} 

  Our current understanding of the Universe is strongly supported by observational facts coming from both studies on the Large Scale Structure via galaxy surveys and from precise CMB measurements. The latter are provided by several sub-orbital experiments as well as satellite measurements now available from the NASA Wilkinson Anisotropy Probe ($WMAP$\cite{WMAP1}) and soon from the ESA Planck mission. The overall picture of a spatially flat Universe is emerging quite strongly, however, a number of fundamental questions are still open, such as the physical nature of Dark Matter and Dark Energy. \par
The Inflationary paradigm provides a physical explanation for the existence of primordial density fluctuations, responsible for structure formation via gravitational instability. While the physics of inflation itself remains somewhat mysterious, the theory is able to make some specific prediction, in particular, it foresees the presence of a stochastic background of gravitational waves. The amplitude and shape of their spectrum is directly related to the inflation energy scale, which is not traced by density fluctuations. The new generation of CMB experiments is reaching a level of observing power that is capable of probing such hypothesis through high precision measurements of CMB anisotropies in temperature and polarization.\par
CMB polarization measurements, in particular, represent a unique tool  to probe energy scales as high as $10^{16}$ GeV, out of reach of any particle accelerator. Specifically, CMB polarization is caused by both density perturbations and gravitational waves. The vector-like nature of polarization is decomposed into orthogonal modes: the curl-free ($E$-modes) and the curl ($B$-modes) components, the latter being generated by gravitational waves. The B-modes amplitude is usually parametrized by the tensor-to-scalar ratio $r$, i.e. the amplitude normalized relative to the known density fluctuations. A tensor amplitude of $r\approx 0.01$ would indicate tremendously high energy scales, comparable to those predicted by Grand Unified Theories (GUTs). Lower values, instead, would point towards breaking of supersymmetry.

\begin{figure}
  \centering
  \includegraphics[height=8cm]{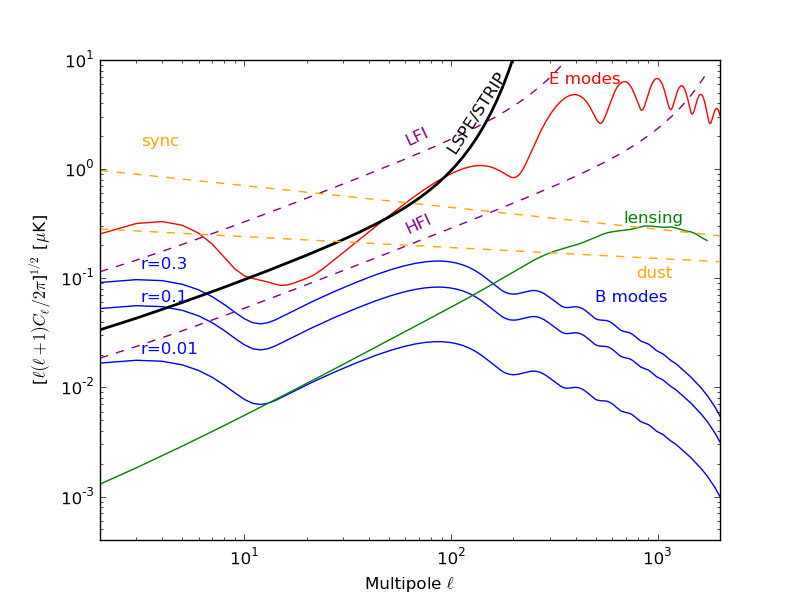}
  \caption{Overall polarized power spectra from CMB (both $E$ and $B$ modes), from galactic foregrounds (synchrotron and dust scaled at 40~GHz) and the expected white-noise sensitivity from LSPE/STRIP compared to both Planck-LFI\cite{mennella2011} and Planck-HFI\cite{HFI}.} 
\label{fig:spectrafull_lspe} 
\end{figure} 

In Fig.~\ref{fig:spectrafull_lspe} we report the level of CMB polarized angular power spectra for the $E$-modes
as well as for the expected amplitude of $B$-modes for different values of $r$, together with noise levels for the Planck instruments and that expected from LSPE/STRIP. The figure also shows the mean levels of contamination from polarized foreground (synchrotron and dust), as well as from lensing, before any attempt of component separation. It is clear  that especially synchrotron emission represents a challenging problem and a
careful selection of observing regions as well as component separation algorithm are required in order to unveil the genuine cosmological signal. The significant improvement of LSPE/STRIP over the expected Planck data at LFI frequencies implies a step forward in our understanding of the microwave sky at large angular scales, where the signature from inflation through the B-modes is most favorable.

 \section{SCIENTIFIC REQUIREMENTS}
 \label{sec:3}

In this section we briefly overview the STRIP main scientific requirements in terms of sensitivity, angular resolution, scanning strategy, calibration accuracy and rejection of the main systematic effects.

\subsection{Sensitivity and angular resolution}
\label{sec_sensitivity_angular_resolution}

  LSPE/STRIP aims at polarization measurements at 43\,GHz with a sensitivity of $\delta Q(U)_{\rm pix} \lesssim 1.8\, \mu {\rm K}$ at 43\,GHz on 1.5$^\circ$ pixels, corresponding to an improvement of a factor $\sim 2.2$ over the Planck-LFI 44 GHz sensitivity after 30 months on the same pixel size\footnote{The comparison has been made using the in-flight measured sensitivity of Planck-LFI reported in Ref.~\citenum{mennella2011}}. 
  Assuming a sky coverage of the order of $\sim 20\%$, a flight duration of two weeks and an angular resolution of 1.5$^\circ$ this sensitivity can be reached in Q-band by an array of 49 coherent polarimeters with a noise temperature of $\sim20$~K. 

  The STRIP focal plane will also host a small array ($\sim 7$ modules) of 90\,GHz polarimeters for cross-checking first-order systematic effects which might be common with the 90\,GHz bolometric channel and assess the contribution of the atmosphere to the measured signal. The final number of 90\,GHz polarimeters will be decided on the basis of the final focal plane design and power budget. Assuming 7 polarimeters with 50\,K noise temperature we will obtain a final $Q/U$ sensitivity of $\sim 7\,\mu$K per 1.5$^\circ$ pixel. 

  Tab.~\ref{tab_strip_performance_requirements} provides a summary of the main STRIP performance requirements.
  
  \begin{table}[h!]
    \caption{STRIP performance requirements}
    \label{tab_strip_performance_requirements}
    \begin{center}
      \begin{tabular}{p{6cm} c c}
	\hline
	\hline
	& 43\,GHz & 90\,GHz\\
	\hline
	Angular resolution ($^\circ$) & 1.5 & 0.7\\
	Observed sky fraction (\%) & \multicolumn{2}{c}{18}\\
	Flight duration (days) & \multicolumn{2}{c}{14} \\
	Noise temperature (including telescope and window -- K) & 27 & 49\\
	Contribution from telescope and window to noise temperature (K) & 7 & 8\\
	Number of polarimeters & 49 & 7\\
	$Q/U$ 1-second sensitivity $(\mu\mathrm{K_{\rm CMB}\,s^{1/2}})$ & 239 & 338 \\
	\hline
      \end{tabular}
    \end{center}
  \end{table}

\subsection{Calibration accuracy and systematic effects budget}
\label{sec_calibration_systematics}

 The STRIP polarimeters are based on the receiver design developed for the QUIET (the Q/U Imaging ExperimenT) ground-based experiment\cite{cleary_2010}. The QUIET polarimeter architecture is very effective in providing direct and stable measurements of the $Q$ and $U$ Stokes parameters with a high level of rejection of systematic effects. This comes at the cost of reduced sensitivity to the total intensity, which is detected in total power although modulated by the 3\,Hz spin frequency and 100\,Hz sampling frequency. The higher 1/$f$ noise in total intensity measurements will likely limit the availability of calibration sources in the sky essentially to the Crab Nebula (for photometric calibration, polarization angle and beam reconstruction) and the Moon (for polarization angle \cite{quiet_2011}). The Moon emission could be strong enough to be detected in total power intensity and be used as a relative photometric calibration source.
  
  The limited availability of polarized calibration sources in the sky calls for trade-offs in the scanning strategy to accommodate the needs of scanning the sky as smoothly as possible with negligible sidelobe inputs from the Earth and of observing calibration sources as often as possible with the largest number of feeds in the focal plane. This trade-off is discussed in Sect.~\ref{sec_scanning_coverage}.

  Limiting the uncertainty due to systematic effects is considered the most important and difficult task in high sensitivity measurements of the CMB polarization. A list of the main expected contributions from systematic effects in STRIP is outlined in Tab.~\ref{tab_main_systematic_effects}, and we require that the overall contribution of systematic effects must not be larger than 10\% of the noise per 1.5$^\circ$ pixel, i.e. $\lesssim \pm 0.1\,\mu$K at Q-band and $\lesssim \pm 0.4\,\mu$K at W-band. 
    
  \begin{table}[h!]
    \caption{Main expected systematic effects in the STRIP instrument}
    \label{tab_main_systematic_effects}
      \begin{center}
	\begin{tabular}{p{10cm}}
	  \hline
	  \hline
	  \multicolumn{1}{c}{\textbf{Optical effects}}\\
	    Beam asymmetry \\
	    Sidelobe pick-up (Earth, payload) \\
	    Spurious-polarization (from telescope, cryostat window, feed-OMT assembly)\\
	  \hline
	  \multicolumn{1}{c}{\textbf{Polarimeter-intrinsic effects}}\\
	    Bandpass mismatch\\
	    Polarimeter imbalance (leakage from total intensity to polarization)\\
	    1/$f$ noise \\
	  \hline
	  \multicolumn{1}{c}{\textbf{Thermal effects}}\\
	    Telescope thermal fluctuations\\
	    Temperature fluctuations in the cryostat\\
	    Temperature fluctuations at the focal plane interfaces\\
	    Temperature fluctuations in the electronics box\\
	    Environmental thermal instabilities (gondola, shields, etc.) \\
	  \hline
	  \multicolumn{1}{c}{\textbf{Effects from back-end electronics}}\\
	    ADC non linearity effects \\
	    DC offset fluctuations \\
	  \hline
	\end{tabular}
      \end{center}

  \end{table}

  Sidelobe pickup will be controlled by a combination of high-performance optical design and shielding in order to obtain $\sim -65$\,dB  rejection level on far sidelobes. Furthermore, the optical interface has been designed around a $-30$\,dB cross-polarization requirement. The QUIET polarimeter architecture has already shown very good balance and low 1/$f$ noise properties \cite{quiet_2011}, so that bandpass mismatch is expected to be the largest contribution to systematic effects coming from the polarimeter itself. Our orthomode transducer (OMT) design is expected to improve the already good match obtained by the QUIET team with their septum polarizer. Other effects of thermal and electrical nature will be efficiently suppressed by the differential nature of the polarimeter and are not expected to be a dominant contribution to the systematic error budget.
  
  A simulation pipeline of the instrument response which includes the systematic effects listed in Tab.~\ref{tab_main_systematic_effects} is currently being developed in order to produce quantitative estimates of the final uncertainties in parallel with the instrument development and testing.

\subsection{Sky coverage and scanning strategy }
\label{sec_scanning_coverage}

  LSPE will map the sky spinning at $\sim 3$ r.p.m. and moving around the North Pole at a latitude around the 78$^{th}$ parallel North  with a constant angular velocity. This scanning strategy will lead to an overall sky coverage of about 20\%, which will be sufficient to reconstruct the angular power spectrum down to multipoles of the order of $\ell\sim 15$. The scanning strategy will be such that the observed sky area will be covered every day by nearly all the STRIP detector beams. This will allow us to carry out null tests that will be key to check the robustness of the results.
  
\begin{figure}[!h]
    \centering
    \includegraphics[height=6cm]{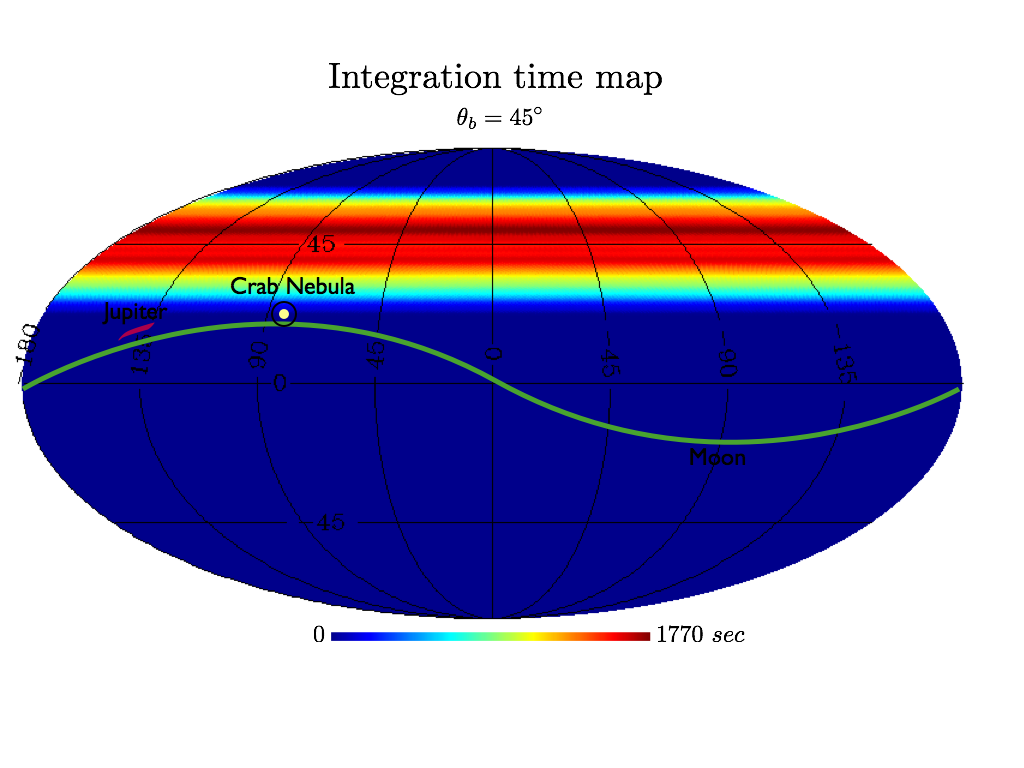}
    \includegraphics[height=6cm]{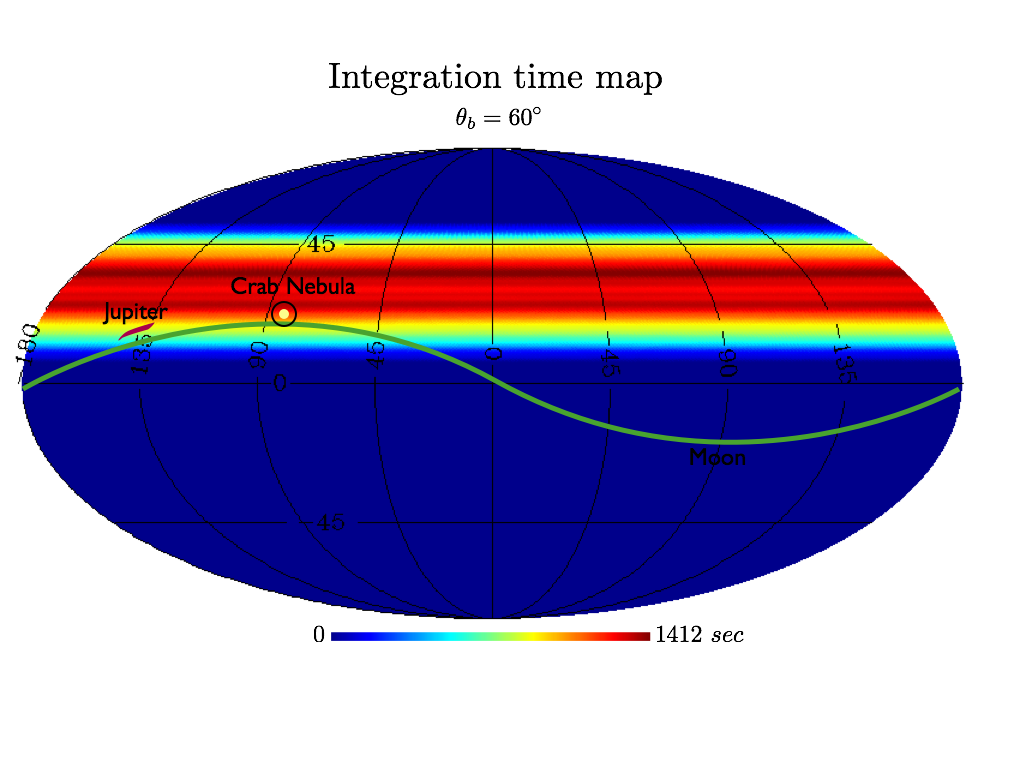}
    \caption{Sky coverage maps (equatorial coordinates) for two different bore-sight angles. On the maps the position of the Crab Nebula, the Moon and Jupiter are plotted. For the last two sources the positions refer to the period between December 2014 and March 2015.} \label{fig:maps} 
    \label{scanning}
  \end{figure}

  An important aspect in the detailed optimization of the final scanning strategy of the STRIP instrument will be the choice of the bore-sight angle $\theta_b$, i.e. the angle between the spin axis and the pointing direction of the telescope: small angles will increase suppression of Earth straylight radiation,  while large $\theta_b$ will result in larger sky coverage. The availability of calibration sources in the focal plane field of view is another key point in this trade-off. In Fig.~\ref{scanning} we show two maps with the pixel integration time for two different bore-sight angles. The maps, in Equatorial coordinates have been generated considering all the antennas in the focal plane with a configuration similar to the one shown in Fig.~\ref{fig:6a_render} for a 15 days flight. The maps also include the positions of the Crab Nebula, the Moon and Jupiter.  With an angle $\theta_b=45^{\circ}$, the sky coverage is $\sim27\%$ but no calibration sources are scanned. On the contrary with ${\theta_b\gtrsim60^{\circ}}$ all the sources would be visible by all the antennas. A detailed scanning plan including periodic changes in $\theta_b$ is under study in order to maximize the observation time of calibration sources with a minimum impact of Earth straylight on scientific CMB observations.

 \section{OVERALL INSTRUMENT CONFIGURATION}
 \label{sec_instrument_configuration}
 
\begin{figure}[h!]
  \begin{center}
      \includegraphics[width=10cm]{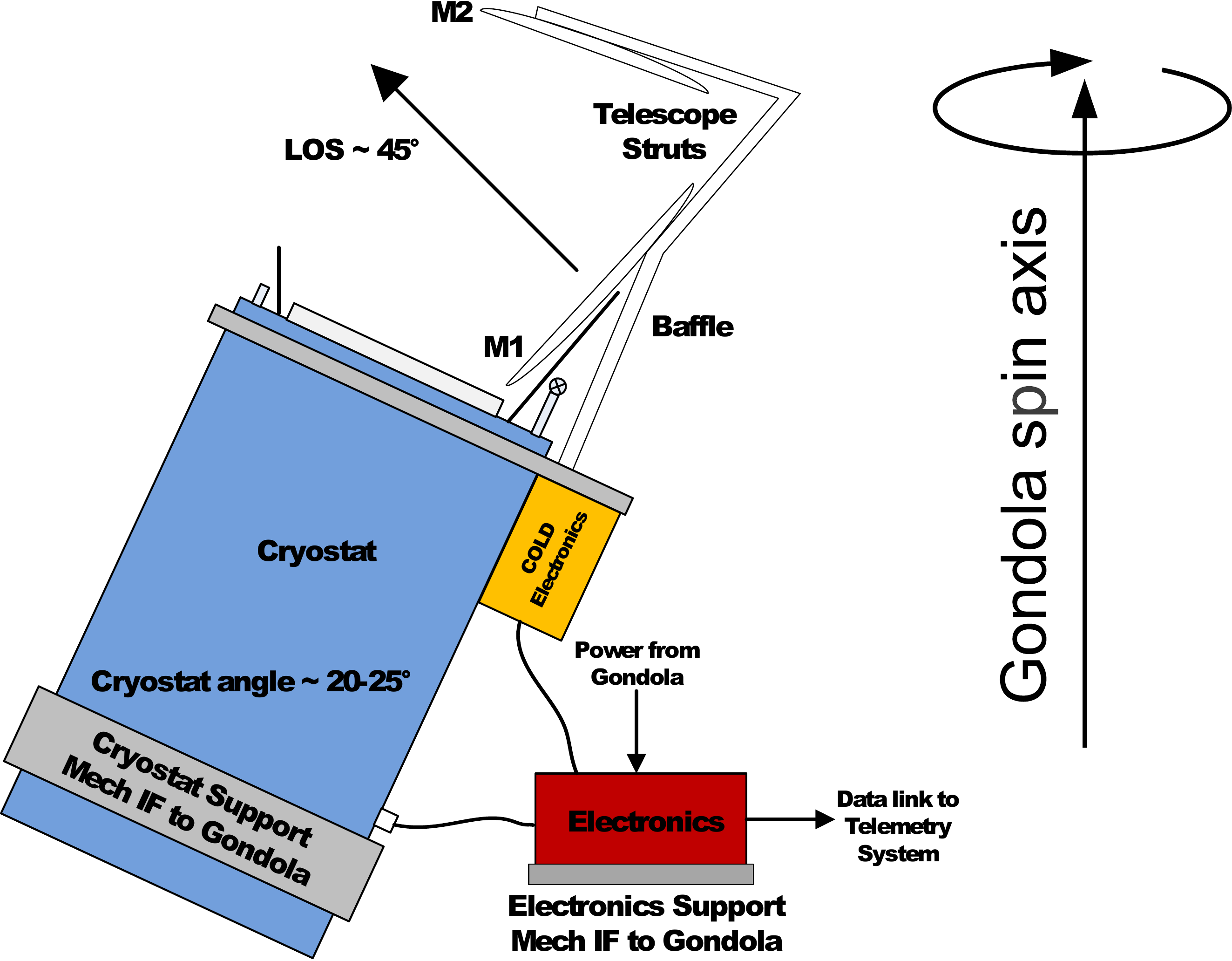}
  \end{center}
    \caption{Schematic of the STRIP instrument. The polarimeter array is enclosed in a 20\,K dewar and observes the sky through a dual-reflector Dragonian side-fed telescope coupled to corrugated feed-horns.}
  \label{fig_strip_overall}
\end{figure}

  The STRIP polarimeter array (see Fig.~\ref{fig_strip_overall}) will be placed on one side of the LSPE gondola and will observe the sky through corrugated feed-horn antennas coupled to a Dragonian side-fed dual reflector telescope with a projected aperture of $\sim 60$\,cm. 

The complete array, constituted by 49 Q-band and 7 W-band polarimeters, will be enclosed in a 20\,K Helium cryostat and the feed-horns array will illuminate the secondary reflector through a low absorption window (see Fig.~\ref{fig:cryo} for a detailed view of the cryostat inner parts). The electronics box with the necessary power suppliers and data acquisition electronics will be placed on the gondola, outside the cryostat, and will be connected to the cold polarimeters via low-parasitic harness and feed-through connectors.

The Q-band array in the STRIP focal plane is constituted by seven modules, each containing seven feed horns arranged in a hexagonal pattern, as shown in Fig.~\ref{fig:6a_render}. The seven W-band polarimeters will be arranged individually around the Q-band focal plane. In the next sections we provide a brief overview of the details of each STRIP subsystem.
 
 \section{Optical system}
 \label{sec:optics}  

\begin{figure}
  \begin{center}
      \includegraphics[height=8cm]{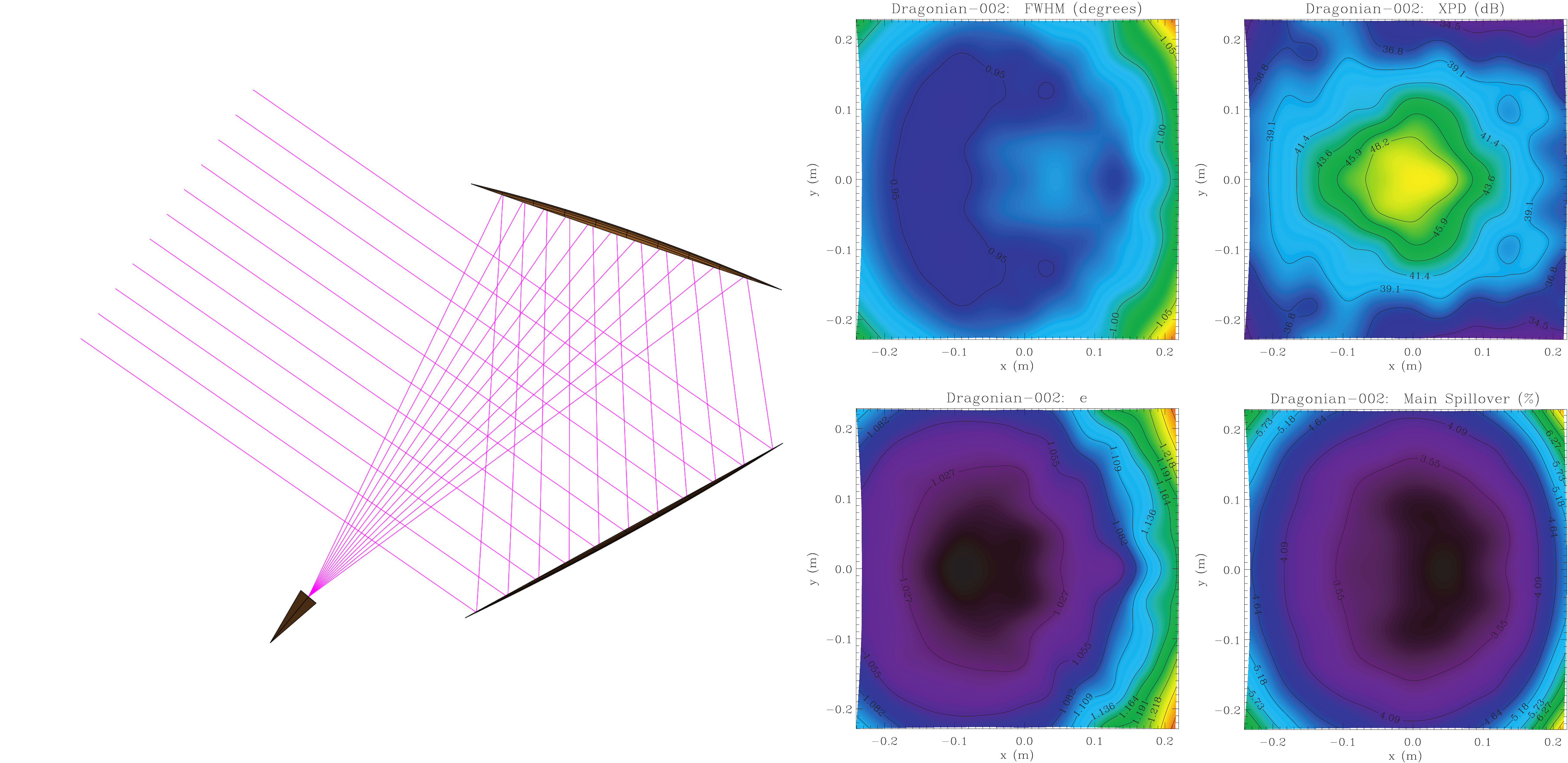}
   \end{center}
   \caption[optics] 
    {\label{fig:optics} 
      In the left panel the STRIP baseline optical scheme is shown. The design has a 60\,cm projected aperture parabolic primary mirror with about 2~m focal length, and a hyperboloid secondary with eccentricity of $-2.40$. A pure $HE_{11}$ hybrid feed with a taper of $-20$\,dB at 15 degree has been considered in the performance simulations. This taper also corresponds to the optics edge taper. The contour plots report (from left to right and top to bottom) the full width half maximum (FWHM) in degree, the cross-polar discrimination in dB, the ellipticity, $e$, and the percentage of power in the main spillover, i.e. power from the feed that scatters the secondary reflector but misses the primary.}
\end{figure}

    The STRIP optical designed is based on the following requirements: (i) a cross-polar discrimination of $-30$\,dB, (ii) an angular resolution of 1.5 degrees, and (iii) a sidelobe rejection level from $-55$\,dB to $-65$\,dB depending on the angular region. A dual reflector telescope has been chosen as the optimal solution in the STRIP frequency range. The selected optical scheme is the compensated side-fed dragonian optics \cite{dragone1978}  which offer excellent performance (angular resolution, polarization purity,  symmetry) over a wide focal region and makes this configuration particularly attractive for CMB polarization measurements\cite{tran2008, tran2010}. The baseline design is reported in Fig.~\ref{fig:optics} together with its main electromagnetic characteristics in the focal region. 

The main challenge imposed by this choice comes from the need to design a fast optics, which for this configuration is in practice limited by $F\#\simeq 2$\footnote{The $F\#$ is the ratio between the focal length of a telescope and its aperture diameter.}. As a consequence, a hard under-illumination  of the mirrors would require focal plane antennas with quite large apertures. In our design, we set an edge taper of -20 dB, which meets the sidelobe requirements, leading to a feed aperture $\sim 50$ mm. This corresponds almost exactly to the projected size of the radiometer's modules, and therefore it does not impact the inter-axis distance and the overall array design.

The feed design is based on conical corrugated horns that are the most performing antennas for such kind of applications\cite{granet2005}. The detailed STRIP optical design will be optimized to accommodate the STRIP focal plane array, to improve straylight rejection and to simplify the mechanical interfaces with the cryostat. This optimization will involve both the telescope and the corrugated profiled antenna design; in particular, several feed horn profiles (linear, sinusoidal, tangential, exponential, hyperbolic) are under evaluation. The choice of the corrugation profile is an additional degree of freedom in the optimization process that will be used to ensure full compliance with the STRIP optical requirements.


 \section{Array design}
 
  \subsection{Overall structure}
\label{sec_structure}
\begin{figure}[!t] 
   \begin{center}
    \includegraphics[height=7cm]{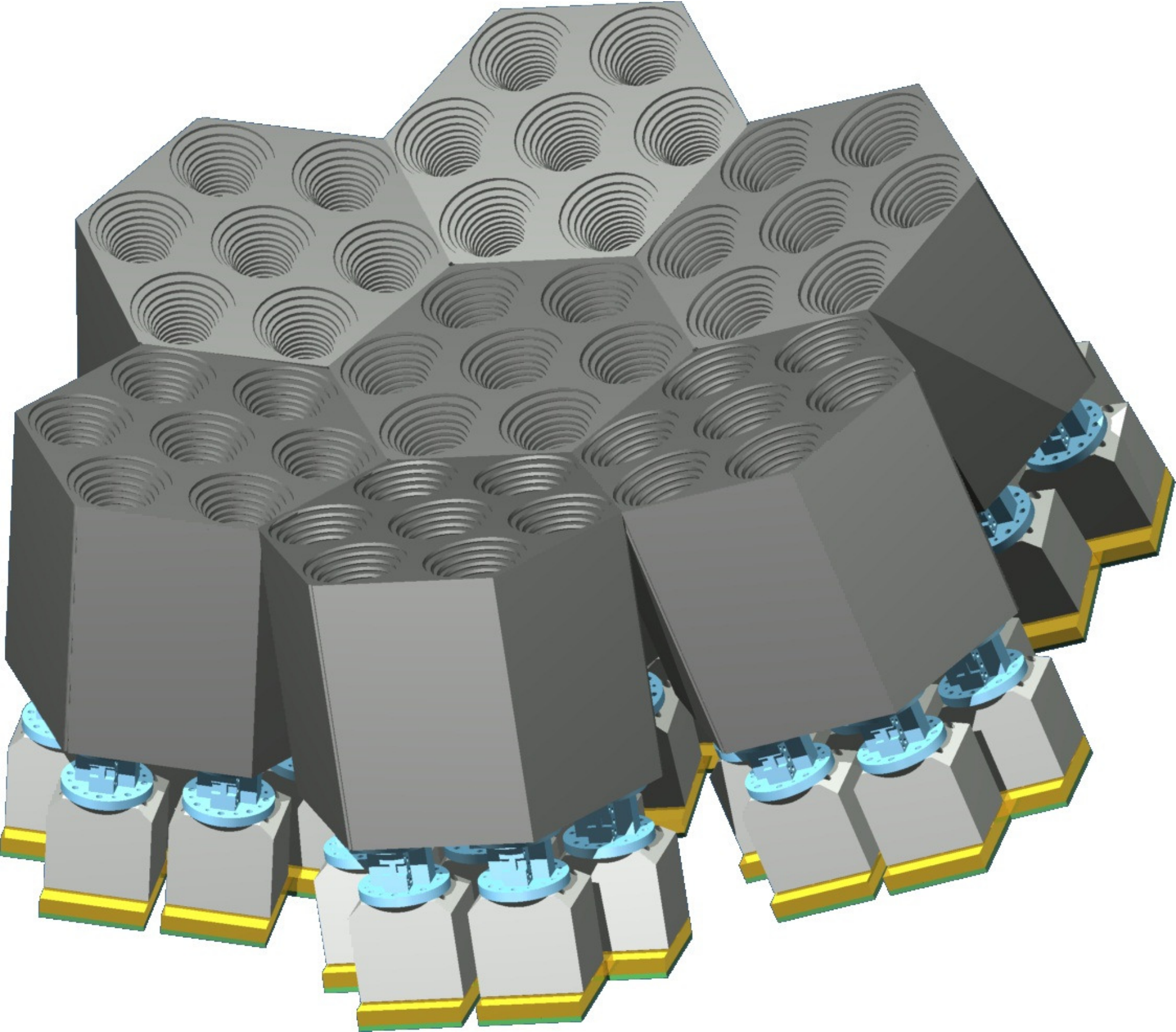}
   \end{center}
   \caption{   \label{fig:6a_render}
    Rendering of the LSPE-STRIP Q-band focal plane unit. Each hexagonal Feed-Horns module is connected to Polarizers (blue), OMTs (gray) and Correlation Units (yellow).}
  \end{figure}

Fig. \ref{fig:6a_render} shows the Q-band array of the Focal Plane Unit (FPU) without supporting frame. The FPU is composed by a 49 element Q-band array at the center of the focal plane, surrounded by seven single W-band feed-horns. The Q-band array is composed by 7 hexagonal modules each including seven feed-horns displaced on an planar honeycomb geometry allowing for maximum compactness. 

This design allows to fit the focal plane surface of the optical system and to minimize handling, testing, and fabrication criticalities being a good compromise between realizing the whole FPU out of 49 single elements or as a unique array.

The FPU will be supported by a single Aluminium frame, thermally an mechanically connected to the cold flange of the cryostat by Aluminium ``pillars''. Being most of the FPU mass concentrated in the feed-horn modules, FPU will be connected to the frame structure trough the hexagonal modules base plate.
  
  \subsection{Feed horns}
\label{sec:horns}
  
The STRIP feed-horns modules will be realized with the platelet technique, consisting in the overlapping of metal plates with holes of proper sizes in order to realize the electromagnetic profile of the feed-horns.

Fabrication techniques normally used for corrugated feed-horns, such as direct machining or electroforming, become time-consuming and cost-ineffective for the production of arrays with large number of elements. The platelet technique allows for reduced times and costs with respect to typical techniques, realizing more elements at the same time ensuring adequate precisions, resulting suitable for mass production \cite{kangas2005,deltorto2011}.

Platelet technique applied to feed-horn manufacturing allows to easily realize any kind of optimized corrugated profile, as an entire corrugation (tooth and groove) can be milled into each plate. Fabrication times can be reduced by choosing the corrugation step to fit market standard platelet thickness, without limiting the electromagnetic design optimization.

This technique requires particular care in plates alignment and in the control of mass and thermal behavior at cryogenic temperatures. In our design the plates are assembled by means of screws located among the feed-horns and tightened directly into the last plate for maximum compactness. Excess material will be removed from each plate and their precise alignment is achieved with external fixtures for lightweight. Plates and screws are made of Aluminium alloys that allow reliable performance in cryogenic conditions and meet the lightweight requirement.



  \subsection{Dual circular-polarization RF chain}
\label{sec_omt}
Each polarimeter is fed by a chain composed by a feed-horn and a circular polarization assembly that splits the incoming radiation into two linearly polarized components proportional to $(E_x+ i\,E_y)/\sqrt{2}$ and $(E_x - i\,E_y)/\sqrt{2}$, where $E_x$ and $E_y$ represent the two linear polarizations of the sky signal.

The dual circular-polarization RF chain designed for the STRIP instrument consists of  a grooved polarizer (converting the two circular polarizations in two linear ones), a turnstile-junction orthomode transducer (routing the two polarized signals to two different rectangular waveguides) and a waveguide circuitry. The latter provides the mechanical mating of the OMT flanges with those of the  polarimetry units.

The selected architecture provides the best trade-off between operating bandwidth (20$\%$) and  electrical performance. The overall design results in a planar symmetric configuration that can be manufactured by combining a multilayer layout with an electrical discharge machining.  Hence, several identical devices can be simultaneously built out of metallic layers of standard thickness (readily available on the market) during the same manufacturing process (cost-effective approach). 

This solution has been chosen based on its high  return loss $(> 30 \rm\, dB)$, isolation $(>  40 \rm\,dB)$ and phase-delay equalization between the two channels and low levels of common and differential insertion loss, cross-polarization $(< - 35 \rm\,dB)$\cite{Cortiglioni01}. All these performance requirements are key to minimize the leakage from total intensity to polarization that would limit the measurement accuracy.

A measurement campaign in the Q-band is underway in order to select the most appropriate metallic layers in terms of machining stability and electrical performances (i.e. insertion losses).

  \subsection{Polarimeters}
\label{sec_polarimeters}

  In the STRIP polarimeters the signals coming from the polarizer assembly are first amplified by two Indium Phosphide (InP) cryogenic low noise amplifiers (LNAs). Then the two signals are phase shifted alternatively by 0 and 180$^\circ$, one at the frequency of $\sim 4$\,kHz and the second at a much lower frequency, $\sim 50$\,Hz. After a further amplification the signals are mixed by a 180$^\circ$ hybrid coupler and split by a 3\,dB power splitter. Two outputs are detected by a diode pair (so-called ``$Q$-diodes'') while the other two are detected by another diode pair (so-called ``$U$-diodes'') after being mixed by a 90$^\circ$ hybrid coupler.

 We have developed a  numerical model of the 43\,GHz STRIP polarimeter using QUCS (``Quite Universal Circuit Simulator'', \url{http://qucs.sourceforge.net/}) an open source electronics circuit simulator software already used in the framework of the Planck-LFI project\cite{zonca2009,zonca_2011}. QUCS allows us to simulate the noise behavior of the circuit also when non linear devices are included. 
  
\begin{figure}[h!]
    \includegraphics[width=17cm]{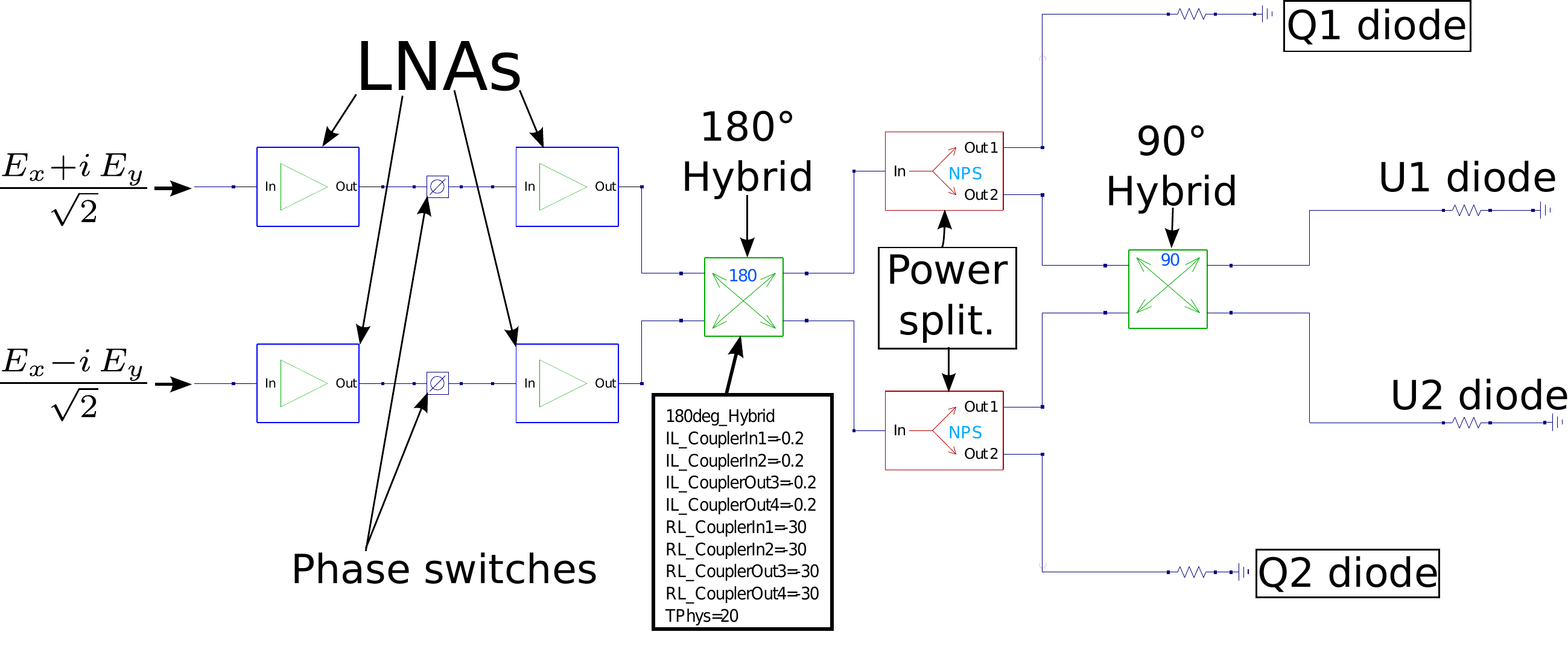}
    \caption{QUCS schematic of the QUIET/STRIP polarimeter. The boxed list shows that each component is defined by a set of parameters; in the figure we show, for simplicity, only the parameters list defining the 180$^\circ$ hybrid coupler.}
    \label{fig_qucs_schematic}
  \end{figure}

  In the QUCS mode, the polarimeter is represented as an electrical network composed by blocks defined by their S-parameter matrices or equations. In general the various blocks have been defined starting from a so-called ``equation defined RF device'', with the exception of the amplifiers and phase shifters where we used the standard QUCS lumped components. The output voltage at each of the four diodes is obtained through an equation which applies the square law to the node voltage. A QUCS schematic of the QUIET polarimeter architecture is shown in Fig.~\ref{fig_qucs_schematic}, which also highlights the main polarimeter components.
  
  We have validated each component by comparing QUCS model outputs with analytic model ones showing a match at a level of 0.0001\%. The next step will be to extend the modeling of the STRIP polarimeter using the expected in-band performance of each component in order to simulate the overall instrument bandpass response in realistic conditions. 
  
  \subsection{Electronics}
\label{sec_electronics}

The STRIP electronic boards will take control of 336 LNAs, 112 phase switches (PS) and 224 detector diodes. Such a large number of signals and conditioning requires a modular architecture to ensure optimal tuning and to fit the tight constraints on power consumption imposed by the long duration night flight.
  
The current design consists of 7 synchronous boards units. Each unit is dedicated to 8 polarimeters and is in charge of the biasing and data acquisition. The units transfer the data to the CPU Unit via Ethernet LAN. Each unit has an in-house designed application-specific integrated circuit (ASIC) chip to bias the RF components of the polarimetric modules, the thirty-two 18\,bit Successive Approximation Analog-to-Digital Converters (SAR ADC) at 1.6 Msps, 1 microcontroller and 1 Field-Programmable Gate Array (FPGA) board for data acquisition and handling. The acquisition from each detector is low noise, fully differential with bias and offset control. The data flow from the ADCs merges into the FPGA which pre-processes the information taking into account the phase switch frequency. Data from each 7-polarimeter module controlled by the same unit are locally stored in a secure digital (SD) card and, for redundancy, the same data are transmitted by the microcontroller to a CPU board housing a RAID1 Solid State Disk (SSD). The pre-amplification section is designed to maximize the gain while minimizing the noise and ensuring optimal dynamic range. A great advantage of an ASIC-based design is the possibility of driving the MMICs (6 LNAs and 2 PSs) both in closed (servo amplifier) or open loop, leaving the possibility to choose between the two configurations even in flight. All the biases are acquired and stored as House-Keeping (HK). The digital section is designed to be ultra low power consuming while maintaining high computing performance.

\section{Thermal design and cryogenic system}
\label{sec_cryogenic}

  The STRIP cryostat is designed to ensure the thermal environment required to operate the instrument for the expected two-weeks flight. Temperature, stability and loads requirements are summarized in Tab.~\ref{tab:cryoreqs}. The instrument baseline design calls for an operating temperature  $\sim20$ K with a temperature stability requirement of 1 mK over a payload spin period (around 1 minute). The active power dissipation (due to the polarimeters load and thermal control)  is expected to be around 1.85 W. The passive loads are mainly due to parasitic heat leaks both conductive (through mechanical struts and harness) and radiative from warmer stages. 

\begin{table}[h]
  \caption{STRIP Main thermal requirements} 
  \label{tab:cryoreqs}
  \begin{center}       
  \begin{tabular}{lcr} 
  \hline
  \hline
  \bf{Requirement} & \bf{Value} & \bf{Comments}  \\
  \hline
  \rule[-1ex]{0pt}{3.5ex}  Operating T &	20 K	& 6~K around this value is an acceptable range for the polarimeters working point   \\
  \rule[-1ex]{0pt}{3.5ex}  Load @20K &	3 W &Including active and passive dissipations  \\
  \rule[-1ex]{0pt}{3.5ex}  T stability &	0.001 K &Peak-to-peak value over a spin period  \\
  \rule[-1ex]{0pt}{3.5ex}  Duration	&15 days	& a 20\% margin (18 days) has been considered for the cryostat design\\
  \hline 
  \end{tabular}
  \end{center}
\end{table}

Limited power resources implied by a night flight prevent the use of mechanical refrigerators and active systems with power consumption higher than few tens of Watts. The best approach, therefore, is to carry cryogens, a choice that has the advantage of being simple and reliable. Three possible solutions have been considered for the 20~K stage required for the polarimeters operating temperature: solid Neon (SNe), liquid hydrogen (LH2) and liquid helium (LHe). Due to the poor availability and extremely high cost of Ne, together with the safety hazards of handling liquid hydrogen (especially in a not-fully-controlled environment, as the launch site for example), LHe has been chosen for the STRIP baseline design.

\begin{figure}
   \begin{center}
   \begin{tabular}{c}
   \includegraphics[height=10cm]{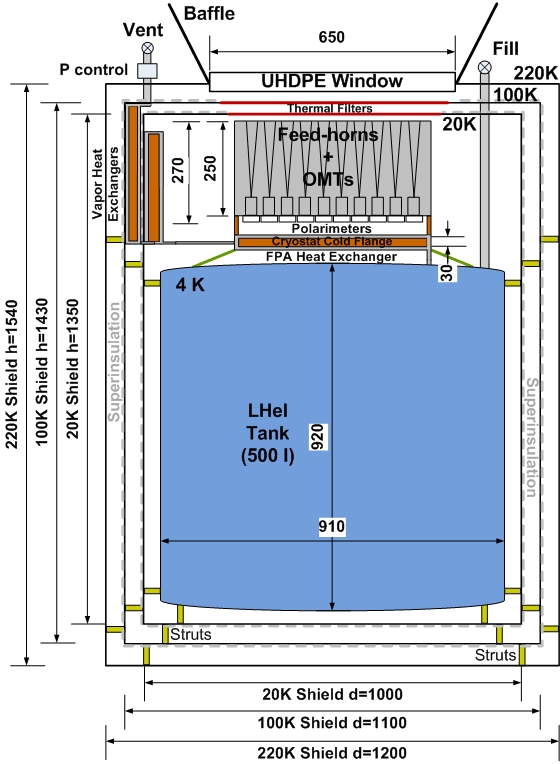}
   \end{tabular}
   \end{center}
   \caption[cryo] 
   { \label{fig:cryo} STRIP cryostat (dimensions in mm)}
\end{figure}

Cooling the array by direct liquid helium evaporation for 15 days would require an unacceptable amount of cryogen volume, of the order of few thousand liters. For this reason, we decided to exploit the high specific heat of evaporating He cold gas to cool the focal plane. By boiling off a calculated mass flow exploiting the parasitic leaks to the LHe tank together with a controlled heater, the cold gas can be circulated in a high-efficiency heat exchanger absorbing the heat from the focal plane and cooling it down to the operating temperature. The gas leaving the 20 K stage is then used to cool the intermediate shield at a temperature $\sim100$ K to intercept parasitic heat leaks. The amount of stored LHe will be of the order of 500 liters to cover the expected mission duration. For this reason, the LHe tank defines the minimum envelope of the cryostat and all other dimensions have been derived from that. The cryostat general configuration and dimensions (in mm) are shown in Fig.~\ref{fig:cryo}. 

One of the main issues of the cryostat design optimization is the minimization of the parasitic heat leaks to colder stages. All main heat transfer paths through conduction (struts, piping, harness) and radiation have been evaluated and reduced by optimizing material selection and mechanical configuration in iterative steps. All feeding and venting pipes will be based on 316L Stainless Steel thin wall tubes while, for the mechanical struts, G10 (or CFRP) will be used.

The large number of electrical connections (around 2000 wires) requires a careful cryoharness  thermal design. Main connections will be realized through very thin copper Flexi cables. Thermal breaks of 20 to 30 cm of Phosphor Bronze Flexi wiring will be used between the stages. These short decoupling stages are enough to reduce harness thermal parasitics, while minimizing the electrical resistance increase along the lines. Radiative heat leaks will be controlled by shielding all stages with MLI Superinsulation.

IR polyethilene or metallic mesh thermal filters prevent thermal radiation from reaching the colder stages.  The cryostat main window must have a very high transmittance to microwave radiation (higher than 0.9) without introducing any spurious polarization effect. For the cryostat main window the evaluation process has identified two best candidates: Ultra-High Density Polyethylene (UHDP) and Plastazote foam.

High thermal stability is a critical aspect in the rejection of instrumental systematic effects. Thermal control will be achieved by a combination of passive and active systems. The passive control exploits thermal masses and resistances of the components (struts and flanges) to damp temperature oscillations during their propagation from the instability source (the cryostat cold flange) to the detectors. The active control is achieved by a closed loop PID-type controller on the focal plane support structure. Since active control on the detectors thermal stage is critical for instrument performance, a system is proposed with two fully redundant heater-sensor pairs.
 
To support the thermal design optimization we have built a reduced model of the focal plane in ESATAN-TMS, assuming the expected properties and characteristics. Steady state simulations have been run to verify equilibrium temperature of the focal plane 
components, while a transient analysis has assessed to check the dynamical response of the FPU with respect to worst case temperature fluctuations at the interface with the cryostat. These simulations are providing an important input to the finalization of the focal plane array thermo-mechanical design and temperature control system.
 
\section{On board processing and communication}
\label{sec_on_board_processing}

  The STRIP electronics instrument will include a CPU module, based on a PC/104 board and a low power consumption Intel Atom processor running a Linux operating system. The communication of the CPU module with the acquisition modules will be performed through LAN ports and a controller-area network (CAN) bus. In particular, LAN communication will be used to receive scientific and housekeeping data from the acquisition modules (MMIC modules), while the CAN bus will be used to send telecommands to the MMICs. The On-Board Software will include four tasks:

\begin{itemize}
  \item{Acquisition}, to write science and HK data received from the acquisition modules on a SSD disk;
  \item{Communications}, to send HK telemetry (TM) using the EGSE (Electrical Ground Support Equipment) for test purposes and to the IRIDIUM based communication; \item{Communications}, to receive telecommands (TC) via the EGSE for test purposes and on the  IRIDIUM; 
  \item{Execution of TC}, to send telecommands to each acquisition unit or sub-system.
\end{itemize}

Attitude and time information will be provided by the attitude control system (ACS), at the frequency of 10 Hz, and could include data from fiber-optic gyroscopes (FOG), the star tracker, a sun sensor, magnetometers and a GPS. Such data will be stored in the CPU SSD and part of it will be included in the essential telemetry.

The quick-look analysis system will be based on KST (\url{http://kst-plot.kde.org/}) with customized modules in order to interface it with the data and the adopted telemetry format. The quick-look will be interfaced to a single DBMS system (Oracle or MySQL are being evaluated) that will contain pre-defined metadata and link to the binary file. The database structure will be defined by using a configurable XML approach in order to achieve maximum flexibility.

The STRIP instrument will produce nearly 120 GB of data during the entire flight operations. Such data size can be safely handled by a single RAID1 mirrored storage system (and a backup unit).

\section{Instrument-level Calibration and testing}
\label{sec_calibration}

  After testing and integration of the STRIP subsystem an Instrument Level test will be performed, with two main objectives: (i) to verify the pre-flight functionality and operation requirements; (ii) to measure instrument performance and asses their compliance with scientific requirements. Furthermore, the test campaign results will provide useful data for checking the consistency with unit level tests and for completing  the missing information needed to  reduce and understand flight data.

Due to  the overall dimensions of STRIP, it is likely that the system level test will be split in two configurations. In the first set-up  the full focal-plane will be integrated in the flight cryostat, with the electronics and the onboard software. This configuration will verify the interface between the cryostat and the focal plane. A reference polarized signal will be fed to the corrugated feed-horns through the cryostat window. This phase will be mainly aimed at investigating the dependence of polarimeters behavior on the cryostat thermal stability and at checking the correct flow of  scientific and housekeeping data with the instrument  in ''flight-like'' conditions. The polarimeters bandwidth will also be measured, by keeping a signal source in the far field of the focal surface to get a plane wave front.

\begin{figure}
  \centering
    \includegraphics[width=4cm]{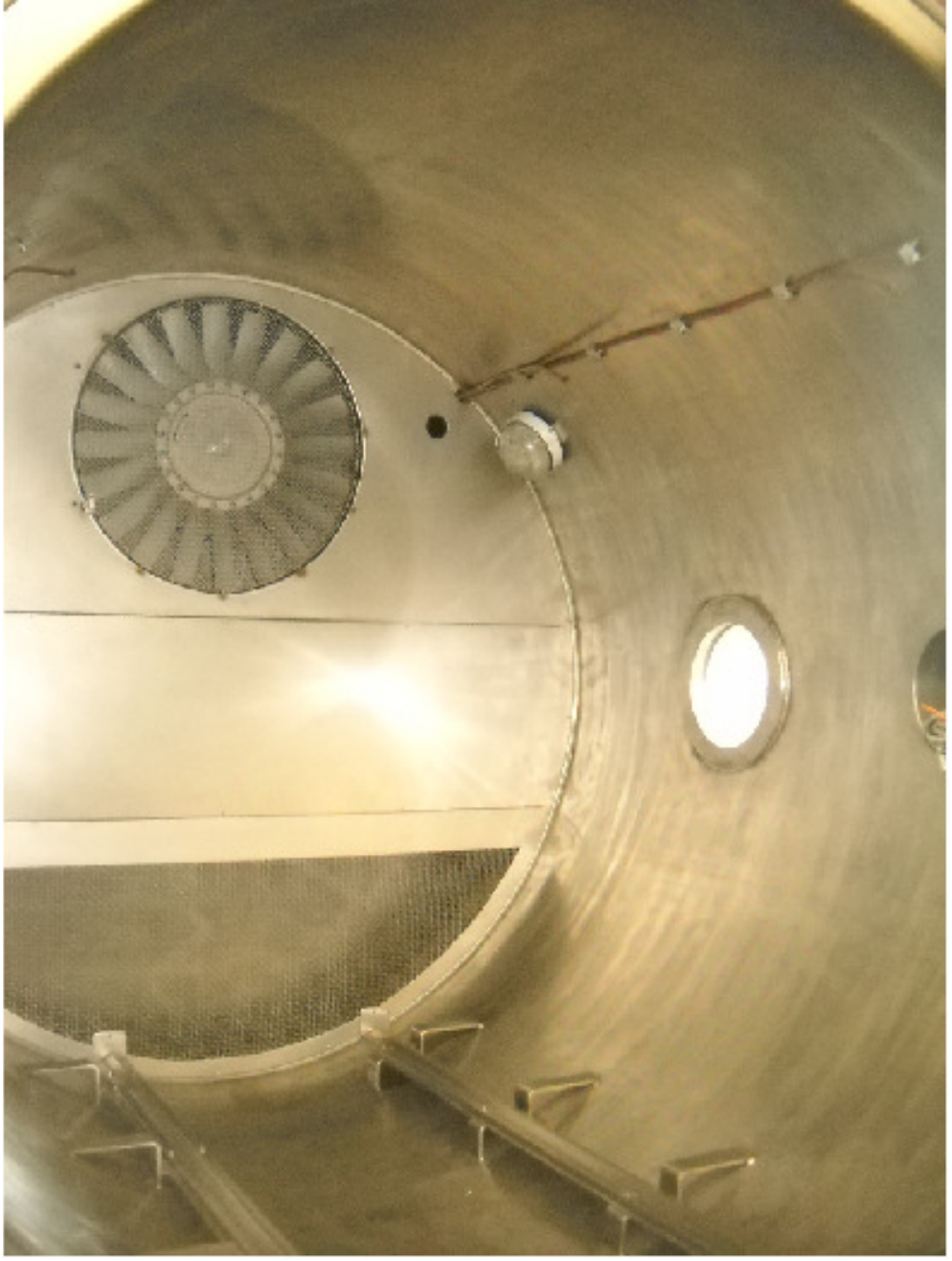}
    \caption {Thermal vacuum chamber available in IASF-Bologna laboratories}
    \label {TEV_chamber}
\end{figure}

The second configuration will verify the  functionality and the overall performance with the flight cryostat replaced by a  mechanical cooler. This set-up minimizes the full system envelope, fitting the IASF-Bologna thermal vacuum chamber (see Fig.~\ref{TEV_chamber}) and allow the system to work for a longer period of testing without stops to refill the tank maximizing schedule efficiency. In this configuration the STRIP polarimeters and the electronics can be coupled to the telescope and a cryogenic polarized calibrator at 4 K: the amplifier bias tuning will be optimized and checked with respect to unit-level results. Linearity response, noise properties, sensitivity, total intensity and Q-U polarization responsivity will be calibrated during this phase, as well as  possible cross talk effects between polarimeters. A basic set of measurements will be repeated at the launch site just before flight: reference signals could be supplied by sources in the far field in order to characterize the full instrument response together with the optics.

\section{Conclusions}

  The Large Scale Polarization Explorer (LSPE) is an ASI-funded balloon experiment designed to measure foreground and CMB polarization on large angular scales with a combined sensitivity of $10\,\mu$K per arcmin. By combining large scale coverage, high sensitivity and an aggressive strategy for foregrounds removal and systematic effects control, LSPE will provide CMB $B$-modes detection with $r\gtrsim 0.001$ and the highest sensitivity detection to date of the $E$-mode reionization signature. 

The LSPE will observe the polarized sky during a long duration arctic night-flight through two instruments: a 20\,K coherent polarimeter array at 43 and 90\,GHz (the STRatospheric Italian Polarimeter, STRIP) and a 0.3\,K bolometric array at 90, 140 and 220\,GHz (the Short Wavelength Instrument for the Polarization Explorer, SWIPE).

The STRIP array is composed by a high sensitivity 49-polarimeter array at 43\,GHz and a 7-polarimeter array at 90\,GHz which will be realized by exploiting the already successful design applied in the QUIET experiment. With a more than a factor 2 improvement in sensitivity compared to Planck-LFI 44\,GHz, the 43\,GHz array will provide state-of-the art measurements of polarized foregrounds and CMB at this frequency band, and will be key in removing the synchrotron and free-free polarized components. The 90\,GHz coherent array will provide measurements overlapping those of SWIPE-90\,GHz and will constitute an important check for common systematic effects. This channel will also provide information about the microwave atmospheric contribution at stratospheric altitudes.

STRIP (and, in general, LSPE) will bring important technological steps forward and will be key for planning future CMB polarization space missions. STRIP, in fact, will be the first large array of coherent microwave detectors ever flown on a balloon while the LSPE constitutes the first case of a night polar long duration balloon flight carrying a payload with two microwave instruments based on different technologies.

\acknowledgments     
 
We gratefully acknowledge support from the Italian Space Agency through contract I-022-11-0 ``LSPE''.



\bibliography{STRIP-SPIE2012}   
\bibliographystyle{spiebib}   

\end{document}